\documentclass[12pt,leqno]{article}
\usepackage{amsmath,amssymb,amsthm,amscd}
\newcommand{\C}{\mathbb{C}}
\newcommand{\Z}{\mathbb{Z}}

\newcommand{\N}{\mathbb{N}}
\newcommand{\im}{\operatorname{Im}}
\newcommand{\vac}{|I\,\rangle}

\newcommand{\End}{\operatorname{End}}
\newcommand{\Resy}{\underset{y=0}{\operatorname{Res\,}}}

\newcommand{\id}{\operatorname{id}}
\newcommand{\Vir}{\mathcal{V}ir}

\newcommand{\Aut}{\operatorname{Aut}\,}
\newcommand{\Outn}{\operatorname{Out}}
\newcommand{\Out}{\operatorname{Out}\,}
\newcommand{\Autn}{\operatorname{Aut}}
\newcommand{\Inn}{\operatorname{Inn}}
\newcommand{\NO}{\,{\raise0.25em\hbox{$\mathop{\hphantom{\cdot}}%
\limits^{_{\circ}}_{^{\circ}}$}}\,}

\theoremstyle{plain}
 \newtheorem{theorem}{Theorem}[section]
	\newtheorem{corollary}[theorem]{Corollary}
	\newtheorem{lemma}[theorem]{Lemma}
	\newtheorem{proposition}[theorem]{Proposition}

\theoremstyle{definition}

\theoremstyle{remark}
\newtheorem{remark}{Remark}[section]
\newtheorem{note}{Note}[section]

\begin{document}
\begin{center}
\begin{Large}
A Note on Free Bosonic Vertex Algebra\\
and its Conformal Vectors\footnote{
This work was partially supported by Grant--in--Aid for Scientific Research, 
the Ministry of Education, Science and Culture.}
\end{Large}
\vskip 3cm
\begin{large}
Atsushi MATSUO
\end{large}
\vskip 0.5cm
Graduate School of Mathematical Sciences,\\
The University of Tokyo,\\
Komaba, Tokyo 153, Japan\\
\vskip 1cm
\begin{large}
Kiyokazu  NAGATOMO
\end{large}
\vskip 0.5cm
Department of Mathematics, \\Graduate School of Science,
Osaka University, \\Toyonaka, Osaka 560, Japan 
\end{center}
\vskip 3cm
\begin{small}
\textbf{Abstract:} We classify all the Heisenberg and conformal vectors
and determine the full automorphism group of the free bosonic vertex algebra
without gradation. To describe it we introduce a notion of inner automorphisms
of a vertex algebra.
\end{small}

\newpage
\section*{Contents}

\begin{description}
\item[0] Introduction
\item[1] Vertex Algebras
	\begin{description}
	\item[1.1] Borcherds' axioms
	\item[1.2] Local fields and operator product expansion
	\item[1.3]	State--field correspondence
	\item[1.4] Conformal vectors and gradation
 \item[1.5] Automorphisms
	\end{description}
\item[2] Free bosonic vertex algebra
	\begin{description}
	\item[2.1] Heisenberg algebra
	\item[2.2] Fock representation
	\item[2.3]	Free bosonic vertex algebra
	\item[2.4] Conformal vectors and standard gradation
	\end{description}
\item[3] Classification of conformal vectors of free bosonic vertex algebra
	\begin{description}
	\item[3.1] Commutative vectors
	\item[3.2] Classification of Heisenberg and Virasoro vectors
	\item[3.3]	Applications
	\item[3.4] Proof of Theorem \ref{theorem:3.1}	
 \end{description} 
\item[4] Discussion

\end{description}

\newpage
\setcounter{section}{-1}
\section{Introduction}
The notion of vertex algebras and their conformal vectors was invented by
Borcherds in \cite{B86}. It is indeed a mathematical
axiomatization of a certain class of chiral algebras appearing in two
dimensional quantum field theory, and the existence of conformal
vector is nothing but peculiar feature, Virasoro symmetry of the
energy--momentum tensor, of two dimensional conformal field theory
initiated by Belavin--Polyakov--Zamolodchikov \cite{BPZ}.
There are a huge number of studies of such structure both in
mathematics and physics, e.g., \cite{B86},\cite{FLM},
\cite{B92},\cite{G89}, \cite{Dong--Lepowsky}, \cite{Li},\cite{LZ}.

Now let us focus on the simplest nontrivial example of vertex algebras,
the free bosonic vertex algebra, whose underlying vector space is 
the polynomial ring with countably many indeterminates identified with the
Fock space of the Heisenberg algebra of charge $0$.
In spite of simplicity, the vertex algebra structure is already
complicated and involves lots of nontrivial features.

In particular, there are infinitely many choices of conformal vectors depending on
one parameter, giving different conformal theories, called the Feigin--Fuks modules
when regarded as Virasoro modules. Our naive interest is whether they exhaust
all the conformal vectors or not. If we assume that a conformal vector should
belong to
the degree $2$ subspace with respect to the natural gradation,
then it is easily seen that they exhaust, as Lian observed in a general framework
\cite{Lian}. He also determined the
automorphism group that preserves the gradation. However, 
if we abandon such restriction concerning the gradation,
there might be other possibility of conformal vectors and automorphisms.

In this note, we classify all the conformal vectors of the free bosonic vertex
algebra without any restriction on the gradation. 
The key observation is that a commutative vector of the vertex algebra, a vector
$c$ satisfying $[Y(c,y),Y(c,z)] = 0$, is proportional to the vacuum vector.
Such uniqueness of the commutative vectors restricts the possibility of the conformal vectors:
they do not have components of degree greater than two. Then the result is that a conformal
vector $v$ is described as
\[
Y(v,z) = \frac{1}{2}\mu^2 I(z) +\mu\alpha(z) + \frac{1}{2}\NO \alpha(z)\alpha(z)\NO +
\lambda\partial \alpha(z),\quad (\lambda,\mu\in\C),
\]
where $I(z) = \id$ is the identity field and $\alpha(z)$ is the Heisenberg field which
generates the vertex algebra. This shows that a conformal vector is
transformed by an inner automorphism to a unique conformal vector in the degree two subspace.
We also classify all the Heisenberg vectors, and determine the full automorphism group: it is a
semidirect product of the automorphism  group that preserves the gradation, which is isomorphic to
$\Z_2$, and the inner automorphism group isomorphic to the additive group $\C$.

Now, we mention the contents of this note. 
In the first section, we recall
definitions and basic properties of vertex algebras. 
This section includes Borcherds' axioms, local fields,
operator product expansion, Goddard's axioms,
existence theorem, gradation, conformal vectors
and automorphisms.
The terminologies used in this note are slightly different
from the literatures, so we give  these definitions precisely
and refer differences of the meanings. 
In particular, we introduce the notion of inner automorphisms suitable for 
our purpose.

The second section is a summary of the free bosonic vertex algebra, which can also
be viewed as giving examples of the notions described in the first section. We first review
the notion of the Heisenberg algebra and and Heisenberg fields, and describe Wick's
theorem which calculates the operator product of fields constructed from Heisenberg
fields. Next we consider the Fock representations $\mathcal{F}_r, (r\in \C)$, of the
Heisenberg algebra and we provide $\mathcal{F}_0$, the vacuum representation, with a
natural structure of vertex algebra. The operations of vertex algebra $\mathcal{F}_0$
can be computed by Wick's theorem. We also describe the conformal vectors and the
standard gradation.

In section 3, we will state, prove and apply our main results on the
classification of conformal vectors. We first prepare the notion of
commutative vectors of a vertex algebra and state the uniqueness of them
in the case of the free bosonic vertex algebra. We then use this result to
classify Heisenberg, Virasoro, and  conformal vectors of the vertex
algebra. We determine the full automorphism group using this
classification. We also mention the complete reducibility of this vertex
algebra as an $sl_2$--module. We finally prove the above mentioned
uniqueness of commutative vectors by a calculation using Wick's theorem.

Final section is devoted to further discussion on 
general vertex algebras, where our method of classifying
all conformal vectors in the free bosonic vertex algebra is
not valid. In this section, we will point out the
difficulties in the general case and give some examples.

\medskip
Throughout this note, we will always work over the complex number field $\C$.
We denote by $\N$ the set of all non--negative integers.

\section{Vertex Algebras}
In this section, we will summarize the axioms and properties of vertex algebras
to fix the notations and give some basic ideas on them.
Most of the contents given in this section are taken from literatures
(\cite{B86},
\cite{B92},\cite{FLM},\cite{G89},\cite{Li}).
In Subsections 1.1--1.5 we review some definitions and basic results,
while in 1.6 we introduce a notion of inner automorphisms which is a new feature.
For
the detail of the contents of Subsections 1.1--1.3, we refer the reader to
\cite{Matsuo--Nagatomo}, in which we will discuss alternative descriptions and
proofs of fundamental results based on the locality of fields.

\subsection{Borcherds' Axioms}
A vertex algebra, in the formulation by Borcherds \cite{B92},
is a $\C$-vector space $V$ equipped with countably many bilinear
binary operations
\[
\begin{array}{ccl}
V \times V&\longrightarrow &V\\
(a,b)&\longmapsto&a_{(n)}b,\quad (n\in \Z)
\end{array}
\]
and an element $\vac \in V$ satisfying the following axioms:
\newline\noindent
(B1) For any pair of elements $a,b\in V$, there exists an
(positive) integer $n_0$ such that	
\begin{equation}\label{eqn:0}
a_{(n)}b = 0\quad\mbox{for all}\quad n\geq n_0.
\end{equation}
\newline
\noindent
(B2) \textbf{(Borcherds identity)} For each triple of elements
$a,b,c\in V$ and integers $k,\ell,m$,
\begin{equation}\label{eqn:1}
\begin{split}
\sum_{i=0}^\infty \binom{k}{i}&(a_{(m+i)}b)_{(k+\ell -i)}c \\
&=\sum_{i=0}^\infty (-1)^i\binom{m}{i}\left(
a_{(k+m-i)}(b_{(\ell + i)}c) - (-1)^mb_{(\ell + m -i)}(a_{(k+i)}c)\right).
\end{split}
\end{equation}
\newline
\noindent
(B3) For any $a\in V$
\begin{equation}\label{eqn:2}
a_{(n)}\vac = \begin{cases}
0&\quad ( n\geq 0)\\
a&\quad ( n = -1).
\end{cases}
\end{equation}
The element $\vac$ is called the vacuum vector\footnote{Perhaps it is better
to call this the identity vector because it corresponds to the identity operator
under the state--field correspondence (cf. Subsection 1.3).}
and the identity (\ref{eqn:1}) is called the Borcherds
identity.\footnote{This identity is nothing but the Cauchy--Jacobi identity of
Frenkel--Lepowsky--Meurman
[FLM, (8.8.29) and (8.8.41)],  while the special case (\ref{eqn:8})  is due to
Borcherds \cite{B86}; we here follow the terminology in \cite{K96}.}

We define $T\in \End V$ by $Ta = a_{(-2)}\vac$, which we call the \textit{translation}
of the vertex algebra $V$.
From the axioms (B1), (B2) and
(B3), we derive the following
formulae:

\begin{align}
T\vac &= 0,\label{eqn:(3)_0}\\
a_{(n)}\vac &= 
\begin{cases}
0&\quad (n\geq 0)\\
T^{(-n-1)}a&\quad (n \leq -1),
\end{cases}
\label{eqn:(2)'}\\
\vac_{(n)}a &= 
\begin{cases}
0&\quad (n\neq -1)\\
a&\quad (n = -1),
\end{cases}
\label{eqn:3}\\
(Ta)_{(n)}b &= -na_{(n-1)}b,\label{eqn:4}\\
a_{(n)}(Tb)& = T(a_{(n)}b) + na_{(n-1)}b,\label{eqn:5}\\
b_{(n)}a & = (-1)^{n+1}a_{(n)}b 
+ \sum_{i=1}^\infty (-1)^{n+i+1}T^{(i)}(a_{(n+i)}b)\label{eqn:6}
\end{align}
where $T^{(n)}= T^n/n!$.
In particular, combining (\ref{eqn:4}) and (\ref{eqn:5}), we
see that
$T$ is a derivation for all the binary operations. We also
have, as particular cases of Borcherds identity,

\begin{align}
[a_{(k)},b_{(\ell)}]& 
= \sum_{i=0}^\infty
\binom{k}{i}(a_{(i)}b)_{(k+\ell-i)},\label{eqn:7}\\
(a_{(m)}b)_{(\ell)} & = \sum_{i=0}^\infty
(-1)^i\binom{m}{i}(a_{(m-i)}b_{(\ell+i)} - (-1)^m
b_{(\ell+m-i)}a_{(i)}).
\label{eqn:8}
\end{align}

\begin{note}
Originally in \cite{B86}, Borcherds took
(\ref{eqn:0}),(\ref{eqn:(2)'}),(\ref{eqn:3}),
(\ref{eqn:6}) and (\ref{eqn:8})  as the set of axioms of
vertex algebras.
\end{note}
Now let us introduce the generating function
\[
Y(a,z) = \sum_{n\in \Z}a_n z^{-n-1}\in (\End V)[[z,z^{-1}]].
\]
Then the axioms  are
restated as follows:
\newline\noindent
(Y1) For any pair of elements $a, b\in V$, the series
\[
Y(a,z)b = \sum_{n\in\Z} (a_{(n)}b)z^{-n-1}
\]
is a Laurent series with only finitely many terms of negative degree.
\newline\noindent
(Y2)\textbf{(Jacobi--identity)}
\[
\begin{split}
\delta(z,y,x)&Y(Y(a,x)b,z)c \\
&=\delta(x,y,z)Y(a,y)Y(b,z)c + \delta(-x,z,y)Y(b,z)Y(a,y)c
\end{split}
\]
where\footnote{It is denoted by
$x^{-1}\delta(\frac{y-z}{x})$ in \cite{FLM} and subsequent papers.},
\[
\delta(x,y,z) = \sum_{n\in \Z}\sum_{i\in \N}(-1)^i \binom{n}{i}x^{-n-1}
y^{n-i}z^i.
\]
\newline\noindent
(Y3) $Y(a,z)\vac$ is a formal power series in $z$ with the
constant term $a$.

The properties (\ref{eqn:(2)'})--(\ref{eqn:6}) are expressed as
\begin{align}
Y(a,z)\vac &= e^{zT}a,\label{eqn:5'} \\
Y(\vac,z)&=  \id_V,\label{eqn:6'}\\
Y(Ta,z) &= \partial_zY(a,z),\label{eqn:7'}\\
[T,Y(a,z)] &= \partial_z Y(a,z),\label{eqn:8'}\\
Y(a,z)b & = e^{zT}Y(b,-z)a,\label{eqn:9'}
\end{align}
respectively.

\subsection{Local Fields and Operator Product Expansion}
Let $M$ be a $\C$--vector space and $A(z)$ a Laurent series with
coefficients in $\End M$:
\[
A(z) = \sum_{n\in \Z}A_{(n)}z^{-n-1}\in (\End M)[[z,z^{-1}]].
\]
Here each $A_{(n)}$ is called a \textit{Fourier mode} of $A(z)$.
The $A(z)$ is called a \textit{field} on $M$ if the series
\[
A(z)u = \sum_{n\in \Z}(A_{(n)}u)z^{-n-1}
\]
has only finitely many terms of negative degree for any $u\in V$.
If $A(z)$ is a field, then so is 
\[
\partial A(z) = \sum_{n\in\Z}(-n-1)A_{(n)}z^{-n-2}.
\]
We set 
\[
A(z)_+ = \sum_{n\geq 0}A_{(n)}z^{-n-1},\quad A(z)_- = \sum_{n < 0}A_{(n)}z^{-n-1}.
\]
If $A(z)$ is a field and $B(z)$ is another one, then the 
\textit{normally ordered product} defined by
\[
\NO A(z)B(z)\NO = A(z)_{-}B(z) + B(z)A(z)_{+}
\]
gives an element of $(\End M)[[z,z^{-1}]]$, which is also a field.
Further we understand the nested normally ordered product by
\[
\begin{split}
\NO A^1(z)&\cdots A^n(z)\NO \\
&= A^1(z)_{-}\NO A^2(z)\cdots A^n(z)\NO +\NO A^2(z)\cdots A^n(z)\NO A^1(z)_{+}.
\end{split}
\]

This construction is generalized as follows:
For each integer $n\in\Z$, define
\[
\begin{split}
A(z)_{(n)}&B(z) = \\
&\Resy A(y)B(z)(y-z)^n|_{|y|>|z|}
- \Resy B(z)A(y)(y-z)^n|_{|y|<|z|}
\end{split}
\]
which we call the \textit{residual $n$--th product}
\footnote{
This is simply called the \textit{$n$--th product} in literatures. 
However, in order to distinguish it from the abstract operations
of a vertex algebra, we prefer to add the adjective \textit{residual}.}.
Here the symbols $(y-z)^n|_{|y|>|z|}$ and
$(y-z)^n|_{|z|>|y|}, (n\in \Z)$, denote the
elements of $\C[[y,y^{-1},z,z^{-1}]]$
obtained by expanding the rational function
$(y-z)^n$ into a series convergent in the regions
$|y|>|z|$ and $|z|>|y|$ respectively:
\begin{align*}
(y-z)^n|_{|y|>|z|}&=
\sum_{i=0}^\infty(-1)^i\binom{n}{i}y^{n-i}z^i,\\
(y-z)^n|_{|z|>|y|}&=
\sum_{i=0}^\infty(-1)^{n+i}\binom{n}{i}y^{i}z^{n-i}.
\end{align*}Then we have
\[
A(z)_{(n)}B(z) = \NO \partial^{(-n-1)}A(z)B(z)\NO 
\]
for $n<0$ and in particular $A(z)_{(-1)}B(z) = \NO A(z)B(z)\NO $.
If $A(z)$ and $B(z)$ are fields, then also are $A(z)_{(n)}B(z)$
and we have
\[
\partial A(z)_{(n)}B(z) = -n A(z)_{(n-1)}B(z).
\]

Now two fields $A(z)$ and $B(z)$ are called (mutually) \textit{local}
\footnote{The locality is sometimes called commutativity. The term
commutativity is also used to mean the identity (7) in literature.
The locality in \cite{LZ} is different from ours, while the commutativity there coincides with
our locality. Here we followed the terminology in \cite{Li} and \cite{K96}.} if
there exists an (positive) integer $m$ such that 
\[
(y-z)^mA(y)B(z) = (y-z)^mB(z)A(y),
\]
namely,
\[
\sum_{i=0}^m(-1)^m\binom{m}{i}(A_{(k-i)}B_{(\ell + i)} - B_{(\ell -i)}A_{(k+i)}) = 0
\]
for any $k,\ell\in \Z$ (cf. \cite{Li}). In this case we also say that $A(z)$ is local with $B(z)$.
If $A(z)$ and
$B(z)$ are local, then so are
$\partial A(z)$ and $B(z)$.  Further, if  $A(z), B(z)$ and $C(z)$ are pairwise local,
then $\NO A(z)B(z)\NO $ and $C(z)$ are local (cf. [Li]). 

If $A(z)$ and $B(z)$ are local, then 
\[
A(z)_{(m)}B(z) = 0\quad \text{for sufficiently large}\quad m,
\]
and we have
\[
A(y)B(z) = \sum_{j=0}^{m-1} 
\left. \frac{A(z)_{(j)}B(z)}{(y-z)^{j+1}}\right|_{|y|>|z|}
+ \NO A(y)B(z)\NO 
\]
and
\[
B(z)A(y) = \sum_{j=0}^{m-1} \left. \frac{A(z)_{(j)}B(z)}{(y-z)^{j+1}}\right|_{|y|<|z|}
+ \NO A(y)B(z)\NO .
\]
Here the normally ordered product $\NO A(y)B(z)\NO $ is defined by
\[
\NO A(y)B(z)\NO  \, = A(y)_{-}B(z) + B(z)A(y)_{+}.
\]

We abbreviate them into a single expression
\[
A(y)B(z)\sim \sum_{j=0}^{m-1}\frac{A(z)_{(j)}B(z)}{(y-z)^{j+1}}.
\]
which is called the \textit{operator product expansion
}\,(OPE). It has sufficient information as far as commutation relations
of Fourier modes are concerned. 

In general we understand the symbol $\sim$
as the equivalence relation in $(\End M)[[y,y^{-1},z,z^{-1}]]$ defined as
\[
K(y,z)\sim L(y,z)
\]
if and only if $K(y,z)u - L(y,z)u$ has only finitely many terms of negative degree in
$y$ and $z$ for any $u\in M$. Then we have
\begin{equation}\label{eqn:a1}
\frac{\NO A(y)B(z)\NO }{(y-z)^n}\sim \sum_{i=0}^{n-1}
\frac{\NO \partial^{(i)}A(z)B(z)\NO }{(y-z)^{n-i}}.
\end{equation}
which holds for expansion both in the region
$|y|>|z|$ and $|y|<|z|$. This is understood to be obtained by
formal substitution of Taylor's formula:
\[
A(y) = \sum_{i=0}^\infty \partial^{(i)}A(z)(y-z)^i
\]
though this does not hold as an equality of formal Laurent series.

We close this subsection with the following fundamental result due to Li.
\begin{theorem}[\cite{Li}]
Let $M$ be a vector space and $\mathcal{O}$ a vector space
consisting of fields on $M$ which are pairwise local. If
$\mathcal{O}$ is closed under residual products and contains the
identity field, then the residual products gives $\mathcal{O}$ a
structure of vertex algebra with the vacuum vector $I(z)$.
\end{theorem}

Li showed this theorem by establishing that the map
\[
\begin{array}{ccc}
Y:&\mathcal{O}\longrightarrow&(\End\mathcal{O})[[\zeta,\zeta^{-1}]]\\
&A(z)\longmapsto&\sum_{n\in\Z}A(z)_{(n)}\zeta^{-n-1}
\end{array}
\]
satisfies Goddard's axioms, which we will explain in the next subsection, with $V = \mathcal{O},
\vac = I(z)$ and
$T = \partial_{z}$. See \cite{Matsuo--Nagatomo} for an alternative
proof. Note that the vector space $M$ need not have a natural structure of
vertex algebra.

\subsection{State--Field Correspondence}
Let us consider the case when $M$ is a vertex
algebra
$V$.  For given $a,b\in V$, take $n_0\in N$ such that $a_{(n)}b = 0\quad(\text{for
all}\quad n\geq n_0)$. Then, letting $m = n_0$, we see the Borcherds identity
immediately implies that
$Y(a,z)$ and $Y(b,z)$ are local (cf. [DL, (7.24)]).
On the other hand, we have property (\ref{eqn:7'}):
\[
\partial_z Y(a,z) = Y(Ta,z)
\]
and a particular case of (\ref{eqn:8}):
\begin{equation}\label{eqn:102}
Y(a_{(-1)}b,z) = \NO Y(a,z)Y(b,z)\NO .
\end{equation}
Therefore the vector space
\[
\mathcal{V} = \left\{Y(a,z)\in (\End V)[[z,z^{-1}]]\,|\,a\in V\right\}
\]
is a set of pairwise  local fields closed under taking the derivative 
and the normally ordered product. From axiom (B2), it follows that the space
$\mathcal{V}$ is naturally isomorphic to the vector space $V$. This is called
the state--field correspondence. Furthermore, by successive use of 
(\ref{eqn:(2)'}) and
formulas (\ref{eqn:7'}) and (\ref{eqn:102}), we have
\begin{equation}\label{eqn:*}
\begin{split}
Y(a^1_{(-j_1-1)}\cdots & a^n_{(-j_n-1)}\vac,z)\\
&= \NO \partial^{(j_1)}Y(a^1_{(-1)}\vac,z)\cdots
\partial^{(j_n)}Y(a^n_{(-1)}\vac,z)\NO 
\end{split}
\end{equation}
where $j_1,j_2,\dots,j_n\in\N$.

Now let us consider the formula (\ref{eqn:7}). This is equivalent to the OPE
\[
Y(a,y)Y(b,z) \sim \sum_{j=0}^{n_0-1} \frac{Y(a_{(j)}b,z)}{(y-z)^{j+1}}.
\]
On the other hand, the formula (\ref{eqn:8}) is nothing but
\[
Y(a,z)_{(j)}Y(b,z) = Y(a_{(j)}b,z).
\]
This means that the vector space
$\mathcal{V}$
equipped with the residual $n$--th product is a vertex algebra with the
vacuum vector being the identity operator, such that the map
\[
\begin{array}{rcc}
Y:V&\longrightarrow&\mathcal{V}\\
a&\longmapsto&Y(a,z)
\end{array}
\]
is an isomorphism of vertex algebras. This is  another version of state--field
correspondence.

In the spirit of the locality, there is another characterization of 
vertex algebras essentially given by Goddard\cite{G89}:
A vertex algebra is a vector space equipped with a linear map
\[
\begin{array}{ccc}
V&\longrightarrow&(\End V)[[z,z^{-1}]]\\
a&\longmapsto&Y(a,z)
\end{array}
\]
whose image is a set of  local fields, an element $\vac\in V$,
and an endomorphism
\[
T:V\longrightarrow V
\]
satisfying (\ref{eqn:(3)_0}), (\ref{eqn:5'}) and (\ref{eqn:8'}):
\[
T\vac = 0,\quad Y(a,z)\vac = e^{zT}a,\quad [T,Y(a,z)] =\partial_z Y(a,z).
\]
Here we can equivalently
replace (\ref{eqn:5'}) by (\ref{eqn:2}).

For the equivalence of this definition and the one described in Subsection 1.1,
we refer the reader to
\cite{Li}, \cite{K96} and \cite{Matsuo--Nagatomo}.

An application is the following existence theorem due to
Frenkel--Kac--Radul--Wang, which is useful in providing a vertex algebra structure on a 
given vector space.

\begin{theorem}[\cite{FKRW}, see also
\cite{K96}]\label{theorem:1.3}
Let $\{A^\lambda(z)|\,\lambda\in \Lambda\}$ be a set of pairwise local
fields on a vector space $V$ such that $A^\lambda(z)\vac$ does not
have terms of negative degree for a given vector $\vac\in V$. If the
set
\[
\{A_{(-j_1-1)}^{\lambda_1}\cdots
A_{(-j_n-1)}^{\lambda_n}\vac\,|\,
\lambda_i\in \Lambda, j_1,\dots,j_n\in \N\}
\]
spans $V$ and there exists an endomorphism $T\in \End V$ satisfying
\[
[T,A^\lambda(z)] = \partial_z A^\lambda(z),
\]
for all $\lambda\in \Lambda$, then $V$ has
a unique structure of vertex algebra with
vacuum vector $\vac$ such that
\[ 
Y(A^\lambda_{(-1)}\vac,z) = A^\lambda(z).
\]
\end{theorem}
Note that the uniqueness is clear, since by (\ref{eqn:8'}) we must
have
\[
\begin{split}
Y(A^{\lambda_1}_{(-j_1-1)}\cdots & A^{\lambda_n}_{(-j_n-1)}\vac,z)\\
&= \NO \partial^{(j_1)}A^{\lambda_1}(z)\cdots
\partial^{(j_n)}A^{\lambda_n}(z)\NO .
\end{split}
\]
In the situation of the theorem, we say that the vertex algebra $V$ is 
\textit{generated} by $\{A^\lambda(z)|\,\alpha\in \Lambda\}$.

\subsection{Conformal Vectors and Gradation}
Let us recall the definition of the \textit{Virasoro algebra}. It is the Lie algebra
\[
\mathcal Vir = \left(\oplus_{n\in\Z}\C L_n \right) \oplus \C C
\]
with the Lie bracket defined by 
\[
[L_m,L_n] = (m-n)L_{m+n} + \delta_{m+n,0}\frac{m^3 - m}{12} C,\quad [C,L_m] = 0.
\]
A representation of $\mathcal{V}ir$ on which $ C$ acts
by a scalar $c\in \C$ is called a representation of 
\textit{central charge} $c$.

Let $(\pi,M)$ be a representation of $\mathcal{V}ir$ of central charge 
$c$. Consider the series
\[
L(z) = \sum_{n\in\Z}\pi(L_n)z^{-n-2}.
\]
If $L(z)$ is a field, then it is local with itself and has the following OPE:
\[
L(y)L(z)\sim \frac{c/2}{(y-z)^4} + \frac{2L(z)}{(y-z)^2} + \frac{\partial L(z)}{y-z}.
\]
Conversely, a field $L(z)$
with such properties defines a representation of $\mathcal{V}ir$ of 
central charge $c$
on $M$ by its Fourier modes. We call such a field $L(z)$ a \textit{Virasoro field}. 
In the sequel, we omit writing $\pi$ and denote as $L(z) = \sum_{n\in\Z}L_n z^{-n-2}$.

Now consider the case when $M$ is a vertex algebra $V$. 
Suppose that there exists a non--zero vector 
$v\in V$ such that
\[
L(z) = Y(v,z).
\]
Then the OPE above is equivalent to
\[
v_{(n)}v =
\begin{cases}
0&\quad (n\geq 4)\\
(c/2)\vac &\quad (n=3)\\
0&\quad (n=2)\\
2v&\quad (n=1)\\
Tv&\quad (n=0).
\end{cases}
\]
Such a vector $v \in V$ is called a \textit{Virasoro vector} of
the vertex algebra $V$.

\begin{remark}\label{remark:virasoro}
Using the skew symmetry of vertex algebras, we can reduce the condition above to
a weaker one
\[
v_{(n)}v =
\begin{cases}
0&\quad (n\geq 4)\\
(c/2)\vac&\quad (n=3)\\
2v&\quad (n=1).
\end{cases}
\]
\end{remark}

Now, for a vertex algebra $(V,Y,\vac, T)$, a Virasoro vector $v \in V$
is called a \textit{conformal vector}\footnote{
In [FLM] the term Virasoro element is used to mean a conformal vector.
} if $L_0 = v_{(1)}$ is
semi--simple and $L_{-1} = v_{(0)}$ coincides with the translation
$T$ of the vertex algebra $V$. A vertex algebra equipped with a conformal vector
is called \textit{conformal vertex algebra}.

Let $V$ be a conformal vertex algebra. Since $L_0$ is semisimple, we have
a direct sum decomposition
\[
V= \oplus_{\Delta\in\C}V^\Delta,\quad V^\Delta = \{a\in V|\,L_0 a = \Delta
a\}.
\]
Then, by (\ref{eqn:7}) and (\ref{eqn:4}), we have
\[
\begin{split}
L_0(a_{(n)}b) &= v_{(1)}(a_{(n)}b)\\
& =a_{(n)}( v_{(1)}b) + (v_{(0)}a)_{(n+1)}b + (v_{(1)}a)_{(n)}b\\
& = (v_{(1)}a)_{(n)}b + a_{(n)}(v_{(1)}b) + (Ta)_{(n+1)}b\\
& =(L_0a)_{(n)}b + a_{(n)}(L_0b) -(n+1)a_{(n)}b.
\end{split}
\]
In other words, if $a\in V^{\Delta_1}$ and $b\in V^{\Delta_2}$, then
$a_{(n)}b\in V^{\Delta_1+\Delta_2 -n-1}$ for any $n\in \Z$.
We take this property as a definition of a gradation of a vertex
algebra. 
Namely, a \textit{gradation} of a vertex algebra is a direct sum
decomposition
\[
V = \oplus_{\Delta\in\C} V^\Delta
\]
such that 
\[
(V^{\Delta_1})_{(n)}(V^{\Delta_2})\subset V^{\Delta_1+\Delta_2 -n-1}
\]
for any $n\in\Z$.

A vertex algebra equipped with a gradation is called a
\textit{graded vertex algebra}. The above argument shows that a
conformal vertex algebra is naturally graded by the eigenspace
decomposition with respect to $L_0$.
In this case, a vector belonging to $V^\Delta$ is said to have
the \textit{conformal weight} $\Delta$.

\begin{note}
For a graded vertex algebra, the axioms are interpreted in
terms of
``operator valued rational functions"(see \cite{FHL}).
\end{note}

\subsection{Automorphisms}
Let $V$ be a vertex algebra. An automorphism of $V$ is an isomorphism
\[
\sigma:V\longmapsto V
\]
of vector spaces which preserves all the products:
\[
\sigma(a_{(n)}b) =\sigma(a)_{(n)}\sigma(b).
\]
In other words, an isomorphism $\sigma$ is an automorphism if and only if
\[
\sigma Y(a,z)\sigma^{-1} = Y(\sigma(a),z)
\]
holds for any $a\in V$.

The group of all automorphisms of the vertex algebra $V$ is denoted by
$\Aut V$. Each automorphism $\sigma \in \Aut V$ preserves the
vacuum vector since
\begin{align*}
\sigma(\vac) &= \vac_{(-1)}\sigma(\vac)\\
& = \sigma(\sigma^{-1}(\vac)_{(-1)}\vac)\\
& = \sigma(\sigma^{-1}(\vac))= \vac,
\end{align*}
and commutes with the translation $T$ since
\begin{align*}
\sigma(Ta)&= \sigma(a_{(-2)}\vac)\\
&=\sigma(a)_{(-2)}\sigma(\vac)\\
&=\sigma(a)_{(-2)}\vac=T(\sigma(a)).
\end{align*}

Let us introduce a notion of inner automorphisms of a vertex algebra. Before giving
the definition, we prepare the following lemma:

\begin{lemma}\label{lemma:1.4}
Let $a$ be an element of a vertex algebra $V$. 
Then, for a non--negative integer $m$,
$a_{(m)}$ is a derivation for all the products of the vertex algebra
if and only if $a_{(0)} = a_{(1)} = \cdots = a_{(m-1)} = 0$.
\end{lemma}

The proof is easily carried out as follows. 
By Borcherds' commutator formula (\ref{eqn:7}), we have
\[
a_{(m)}(b_{(n)}c) = (a_{(m)}b)_{(n)}c + b_{(n)}(a_{(m)}c) +
\sum_{j=0}^{m-1}\binom{m}{j}(a_{(j)}b)_{(m+n-j)}c.
\]
So if $a_{(0)} = a_{(1)} = \cdots = a_{(m-1)} = 0$, then 
$a_{(m)}$ is a derivation for all the $n$--th products. Conversely, if $a_{(m)}$
is a derivation, we have
\[
\sum_{j=0}^{m-1}\binom{m}{j}(a_{(j)}b)_{(m+n-j)}c = 0.
\]
Putting $c = \vac$ and $n=-2$, we have 
$a_{(m-1)}b = 0$ for any $b\in V$. Inductively we deduce that
$a_{(0)} = a_{(1)} = \cdots = a_{(m-1)} = 0$. We call such a derivation as in the lemma an
\textit{inner derivation}\footnote{ The notion of inner derivation in 
\cite{Lian} coincides with our inner derivation of level $0$.} \textit{of level} $m$.

Let $a$ be an element of $V$ with $a_{(0)} = a_{(1)} = \cdots = a_{(m-1)} = 0$.
If the exponential
\[
\sigma = \exp{(a_{(m)})} = \sum_{n=0}^\infty\frac{a_{(m)}{}^n}{n!}
\]
of the derivation $a_{(m)}$ makes sense, then it give rise to an 
automorphism of $V$. We call such an automorphism an \textit{inner automorphism}
of \textit{level} $m$. We denote
\[
\Inn_{(m)}\, V =\langle
\sigma\in\Aut V\,|\, \sigma =\exp{(a_{(m)})}\quad
\text{for some}\quad a\in V\rangle.
\]
Then, since
\[
a_{(m)} = -\frac{1}{m+1}(Ta)_{(m+1)},\quad (m\geq 0)
\]
we have the following inclusions:
\[
\Inn_{(0)}\,V \subseteq \Inn_{(1)}\,V \subseteq \Inn_{(2)}\,V\subseteq
\cdots.
\]
Let us denote by $\Inn\, V$  the union of $\Inn_{(n)}\,V$.
Note that all $\Inn_{(m)}\, V$ and $\Inn\, V$ are normal subgroups of
$\Aut V$ since
\[
\sigma\exp{(a_{(m)})}\sigma^{-1} = \exp{(\sigma(a)_{(m)})}
\]
for any $\sigma\in \Aut V$.

Now let us turn to the case when $V$ is graded:
$V = \oplus_{\Delta\in\C}V^\Delta$. We denote the subgroup of $\Aut
V$ consisting of all automorphisms that preserve the gradation by
\[
\Autn^0\,V = \left\{
\sigma \in \Aut V\,|\, \sigma(V^\Delta) = V^\Delta\quad \text{for all}
\quad \Delta\in \C\right\}.
\]
We set
\[
\Inn^0_{(m)}\, V = \Inn_{(m)}\, V\cap \Autn^0\, V,\quad \text{and}
\quad
\Inn^0\, V = \Inn\, V\cap \Autn^0\,V.
\]
Since the gradation satisfies
\[
(V^{\Delta_1})_{(m)}(V^{\Delta_2})\subset
V^{\Delta_1+\Delta_2-m-1},
\]
the inner automorphism $\sigma = \exp{(a_{(m)})}$ preserves
the gradation if and only if $a\in V^{m+1}$.
Hence we have
\[
\Inn^0_{(m)}\,V =\langle
\sigma\in \Aut V\,|\,
\sigma = \exp{(a_{(m)})}\quad\text{for some}\quad 
a\in V^{m+1}\rangle.
\]

We define the outer automorphism groups as
\begin{align*}
\Outn_{(m)}\, V &= \Aut V/\Inn_{(m)}\, V,\\
\Out V &= \Aut V/\Inn\, V,\\
\Outn^0_{(m)}\, V &= \Autn^0\, V/\Inn^0_{(m)}\, V,\\
\Outn^0\, V &= \Autn^0\, V/\Inn^0\, V
\end{align*}
though they will not appear in the rest of the paper.

\section{Free Bosonic Vertex Algebra}
This section is devoted to give a definition and fundamental properties of 
the free bosonic vertex algebra, which is the main objective of this article.

\subsection{Heisenberg Algebra}
Let us recall the definition of the Heisenberg algebra. 
Consider the Lie algebra
\[
\mathcal{H} = \left(\oplus_{n\in\Z}\C H_n\right) \oplus \C K
\]
with the Lie bracket defined by
\[
[H_m,H_n] =m\delta_{m+n,0}K,\quad [H_m,K] = 0.
\]

Let $V$ be a representation of $\mathcal{H}$ on which
$K$ acts by a non--zero scalar. Then, by rescaling $H_n$, we may assume that
$K$ acts by $1$. Such a representation is nothing but a module of the
associative algebra $\mathcal{A}$:
\[
\mathcal{A} = U(\mathcal{H})/U(\mathcal{H})(K-1),
\]
where $U(\mathcal{H})$ is the universal enveloping algebra of $\mathcal{H}$.
Let us denote the generators of $\mathcal{A}$
corresponding to $H_n$ by $\alpha_n$. Then $\mathcal{A}$ is the associative algebra generated
by $\{\alpha_n|\,n\in\Z\}$ with
fundamental relations
\[
\alpha_m\alpha_n - \alpha_n\alpha_m = m\delta_{m+n,0}\quad (m,n\in\Z).
\]
The algebra $\mathcal{A}$ is called the \textit{Heisenberg algebra}
\footnote{The term Heisenberg algebra is used to mean the Lie algebra
$\mathcal{H}$ in some literatures.}.

Let $(\pi,M)$ be a representation of $\mathcal{A}$. Consider the series
\[
\alpha(z) = \sum_{n\in \Z}\pi(\alpha_n)z^{-n-1}.
\]
If $\alpha(z)$ is a field, then it is local with itself and has the following OPE
\[
\alpha(y)\alpha(z) \sim \frac{1}{(y-z)^2}.
\]
Conversely, a field with such properties defines a $\mathcal{A}$--module on $M$.
We call such a field a (normalized) \textit{Heisenberg field}. In the sequel, we omit
writing
$\pi$.

\begin{note}
Since the Lie algebra $\mathcal{H}$ is the affinization of the abelian Lie
algebra
$\mathfrak{u}(1)$:
\[
\mathcal{H}\cong \left(\C[t,t^{-1}]\otimes \mathfrak{u}(1)\right) \oplus \C K,
\]
a Heisenberg field is called a $U(1)$ --current in physics.
\end{note}

Now consider the case when $M$ is a vertex algebra $V$. 
Suppose that there exists an element $h \in V$ such that 
\[
\alpha(z) = Y(h,z)
\]
is a Heisenberg field. Then the OPE above is equivalent to
\[
h_{(n)}h = \begin{cases}
0&\quad (n\geq 2)\\
\vac&\quad (n=1)\\
0&\quad (n=0).
\end{cases}
\]
We call such an element $h\in V$ a (normalized) \textit{Heisenberg vector}
of the vertex algebra $V$.

Let $M$ be a vector space and $\alpha(z)$ a Heisenberg field.
Consider the space $\mathcal{O}$ of all the fields 
which are obtained from $\alpha(z)$ and $I(z)$ by successive use of normally ordered
product and derivative. Let us first consider the OPE of fields 
$\partial^{(p)}\alpha(y)$ and $\partial^{(q)}\alpha(z)$ for $p,q\in \N$.
It is given by
\[
\begin{split}
\partial^{(p)}\alpha(y)\partial^{(q)}\alpha(z)&\sim \partial_y^{(p)}\partial_z^{(q)}
\frac{1}{(y-z)^2}\\
& = (-1)^p\frac{(p+q+1)!}{p!q!}\frac{1}{(y-z)^{p+q+2}}.
\end{split}
\]
The RHS is denoted as
\[
\langle \partial^{(p)}\alpha(y)\partial^{(q)}\alpha(z)\rangle
= (-1)^p\frac{(p+q+1)!}{p!q!}\frac{1}{(y-z)^{p+q+2}}
\]
and is called the \textit{contraction}. In particular, this does not vanish for any
$p,q\in \N$.

Now consider the general case: The OPE of two fields in $\mathcal{O}$ is
given by the following theorem, together with the formula (\ref{eqn:a1}).

\begin{theorem}[Wick's theorem]\label{theorem:2.1}
We have
\[
\begin{split}
(\NO \partial^{(p_1)}\alpha(y)&\cdots \partial^{(p_m)}\alpha(y)\NO )
(\NO \partial^{(q_1)}\alpha(z)\cdots \partial^{(q_n)}\alpha(z)\NO ) \\
& = \sum_{d=0}^{\max{(m,n)}}\frac{1}{d!}
\sum_{
\begin{tiny}
\begin{array}{l}
\phi:[1,d]\rightarrow[1,m],\\
\psi:[1,d]\rightarrow[1,n]
\end{array}
\end{tiny}}
\prod_{i=1}^d
\langle \partial^{(p_{\phi(i)})}\alpha(y)\partial^{(q_{\psi(i)})}\alpha(z)\rangle\times\\
&\quad \quad \times \NO \prod_{j\in [1,d]\setminus\im
\phi}\partial^{(p_j)}\alpha(y)
\prod_{j\in [1,d]\setminus\im \psi}\partial^{(q_j)}\alpha(z)\NO 
\end{split}
\]
where the second summation is over all injective maps $\phi$ and $\psi$.
\end{theorem}
For example, we have
\begin{equation}\label{WickFormula}
\begin{split}
&(\NO \alpha(y)\alpha(y)\NO )(\NO \alpha(z)\alpha(z)\NO )\\
&=\NO \alpha(y)\alpha(y)\alpha(z)\alpha(z)\NO 
+4\frac{1}{(y-z)^2}\NO \alpha(y)\alpha(z)\NO 
+2\frac{1}{(y-z)^4}\\
&\sim 2\frac{1}{(y-z)^4}+4\frac{1}{(y-z)^2}\NO \alpha(z)\alpha(z)\NO 
+ 4\frac{1}{y-z}\NO \partial\alpha(z)\alpha(z)\NO ,\\
&\\
&\partial\alpha(y)(\NO \alpha(z)\alpha(z)\NO )\\
&=
\NO \partial\alpha(y)\alpha(z)\alpha(z)\NO 
-4\frac{1}{(y-z)^3}\alpha(z)\sim -4\frac{1}{(y-z)^3}\alpha(z),\\
&\\
&(\NO \alpha(y)\alpha(y)\NO )\partial\alpha(z)\\
&= 
\NO \alpha(y)\alpha(y)\partial\alpha(z)\NO 
+4\frac{1}{(y-z)^3}\alpha(y)\\
&\sim 4\frac{1}{(y-z)^3}\alpha(z) +4\frac{1}{(y-z)^2}\partial\alpha(z)
+4\frac{1}{(y-z)^2}\partial^{(2)}\alpha(z),\\
&\\
&\partial\alpha(y)\partial\alpha(z)= \NO \partial \alpha(y)\partial\alpha(z)\NO 
-6\frac{1}{(y-z)^4}\sim  -6\frac{1}{(y-z)^4}.
\end{split}
\end{equation}

\subsection{Fock Representation}

Let $\C [x_1,x_2,\dots]$ be the polynomial ring in countably many
variables. We define the \textit{degree} of a polynomial by setting 
\[
\deg (x_{i_1}\cdots x_{i_k}) = i_1 +\cdots+i_k
\]
for a monomial $x_{i_1}\cdots x_{i_k}$.

Now, for each complex number
$r\in\C$, consider the
$\mathcal{A}$--module
\[
\pi_r:\mathcal{A}\longrightarrow \End \mathcal{F}_r,\quad \mathcal{F}_r = \C [x_1,x_2,\dots]
\]
defined by
\[
\alpha_n\longmapsto
\begin{cases}
n\frac{\partial}{\partial x_n}&\quad (n>0)\\
r&\quad (n=0)\\
x_{-n}&\quad (n<0).
\end{cases}
\]
This is an irreducible representation of $\mathcal{A}$ and is called the 
\textit{Fock representation} of \textit{charge} $r$.
The unit $1$ of the ring $\C [x_1,x_2,\dots]$ satisfies
\[
\alpha_n 1 = \begin{cases}
0&\quad (n>0)\\
r&\quad (n=0)\\
x_{-n}&\quad (n<0).
\end{cases}
\]
and the space $\mathcal{F}_r$ is generated by $1$ as an $\mathcal{A}$--module.
For the charge $r=0$, the Fock representation $\mathcal{F}_0$ is called
the \textit{vacuum representation} and plays the special role: $V =\mathcal{F}_0$ has
a natural structure of vertex algebra and $M = \mathcal{F}_r$ are its modules.

\begin{remark}\label{remark:2.1}
Let us give another description of
Fock representation
$\mathcal{F}_r$.
We recall that $\mathcal{A}$ has the triangular decomposition:
\[
\mathcal{A} = \mathcal{A}_{-}\otimes \mathcal{A}_{0}\otimes \mathcal{A}_{+}
\]
where $\mathcal{A}_{-}$, $\mathcal{A}_{0}$ and $\mathcal{A}_{+}$
are  the commutative
subalgebras generated by $\{\alpha_n|\,n<0\}$, $\{\alpha_0\}$ and  
$\{\alpha_n|\,n>0\}$ respectively.
For each $r\in \C$, let $\C_r = \C |r\rangle$ be the 
one dimensional representation of $\mathcal{A}_{0}\otimes \mathcal{A}_{+}$ defined by
\[
\alpha_n|r\rangle =\begin{cases}
0&\quad (n>0)\\
r|r\rangle&\quad (n=0).
\end{cases}
\]
Consider the induced module
\[
\mathcal{A}\otimes_{\mathcal{A}_{0}\otimes \mathcal{A}_{+}}\C_r.
\]
Then this is in fact an irreducible $\mathcal{A}$--module which is isomorphic
to the Fock representation $\mathcal{F}_r$, where $|r\rangle$ is identified 
with the unit $1\in \mathcal{F}_r$.
\end{remark}
\subsection{Free Bosonic Vertex Algebra}
Let $V = \mathcal{F}_0$ be the vacuum representation of 
Heisenberg algebra $\mathcal{A}$. Consider the series
\[
\alpha(z) = \sum_{n\in \Z}\alpha_n z^{-n-1}.
\]
Then this is a Heisenberg field and we have 
\[
\alpha(z)1 = x_1 +\sum_{n\geq 1}x_{n+1} z^n.
\]
Under this situation, let us seek for a vertex algebra structure on $V$
so that we have
\[
\alpha(z) = Y(x_1,z)
\]
and $\vac = 1$. Note that the constants $c\in \C [x_1,x_2,\dots]$ are the only elements which
have the property that $\alpha(z)c$ has no term of negative degree.
Now, such a structure is unique
if it exists, because $V$ is spanned by
\[
S = \{\alpha_{-i_1-1}\cdots\alpha_{-i_n-1}1|\,n\in\N,i_1,i_2,\dots,i_n\in \N\}.
\]
Note that $\alpha_{-i_1-1}\cdots\alpha_{-i_n-1}1= x_{i_1}\cdots x_{i_n}$.
Thus we must have
\begin{equation}\label{eqn:a2}
Y(x_{i_1}\cdots x_{i_n},z) = \NO \partial^{(i_1)}\alpha(z)\cdots
\partial^{(i_n)}\alpha(z)\NO 
\end{equation}
To apply the existence theorem, we define $T$ by
\[
T x_n = nx_{n+1}
\]
on the generators. We can uniquely extend this to a derivation 
of an associative algebra $V$, since
$V = \C[x_1,x_2,\dots]$ is a polynomial ring. Indeed, the operator
\[
T = \sum_{n=1}^\infty nx_{n+1}\frac{\partial}{\partial x_n}
\]
has the desired property. Then it is easy to check that
\[
T\vac = 0,\quad [T,\alpha(z)] = \partial_z \alpha(z)
\]
and the existence theorem \ref{theorem:1.3} shows that
(\ref{eqn:a2}) gives rise to a vertex algebra structure on $V$. We call
this vertex algebra $V = \mathcal{F}_0$ the \textit{free bosonic vertex algebra}.

Another way of constructing the free bosonic vertex algebra is as follows:
Take any $M = \mathcal{F}_r (r\in\C)$ and consider the space $\mathcal{O}$
spanned by all the fields obtained from $\alpha(z)$ and $I(z)$ by successive use
of  normally ordered product and derivative. Then the space 
$\mathcal{O}$ is closed under the residual products so that it has a structure
of vertex algebra by Li's theorem.
Moreover, it has a structure of $\mathcal{A}$--module via 
\[
\begin{array}{cc}
\mathcal{A}\longrightarrow&\End \mathcal{O}\\
\alpha_n\longmapsto&\alpha(z)_{(n)}
\end{array}
\]
and the identity field $I(z)$ satisfies
\[
\alpha(z)_{(n)}I(z) = 0\quad (n\geq 0).
\]
Therefore, by the construction of Remark \ref{remark:2.1},
we have a unique $\mathcal{A}$--module
map
\[
\mathcal{F}_0\longrightarrow \mathcal{O}
\]
which sends $|0\rangle$ to $I(z)$. This map is an isomorphism because
$\mathcal{O}$ is generated by $I(z)$ as an $\mathcal{A}$--module
and $\mathcal{F}_0$ is irreducible.
So we can introduce a structure of vertex algebra on $\mathcal{F}_0$ 
via this isomorphism, which coincides with the one described above.

Now, Wick's theorem \ref{theorem:2.1} and the formula (\ref{eqn:a1}) enable us to calculate the
residual products of fields in $\mathcal{O}$, so that we can calculate 
the binary operations of the
free bosonic vertex algebra through the state--field correspondence.

\subsection{Conformal Vectors and Standard Gradation}
Let $M$ be a vector space and $\alpha(z)$ a Heisenberg field on $M$.
Then, by (\ref{WickFormula}), the field
\[
L(z) = \frac{1}{2}\NO \alpha(z)\alpha(z)\NO  + \lambda\partial \alpha(z),\quad (\lambda \in \C)
\]
satisfies
\[
L(y)L(z)\sim \frac{1/2 - 6\lambda^2}{(y-z)^4}
+ \frac{2L(z)}{(y-z)^2} + \frac{\partial L(z)}{y-z}.
\]
Namely, the field $L(z)$ is a Virasoro field of central charge $1-12\lambda^2$.

\begin{note}
For $M = \mathcal{F}_r$, the $\mathcal{V}$ir--module obtained by
$L(z)$ is called the \textit{Feigin--Fuks} module and possesses an important
position in mathematical physics.
\end{note}

In case when $M$ is the free bosonic vertex algebra $V = \mathcal{F}_0$,
we have
\[
L(z) = Y(\frac{1}{2}x_1^2 +\lambda x_2,z)
\]
and $L_{-1} = T$, where $T$ is the translation of 
the vertex algebra $V$ and 
\[
L_0 = \sum_{n=1}^\infty nx_n\frac{\partial}{\partial x_n},
\]
which is semi--simple. Therefore,
\begin{equation}\label{eqn:conformal}
v = \frac{1}{2}x_1^2 +\lambda x_2,\quad (\lambda \in \C)
\end{equation}
is a conformal vector of $V$. We call the gradation obtained by 
$L_0$ the \textit{standard gradation} of $V$.
The corresponding decomposition is
$V = \oplus_{n=0}^\infty V^n$, where
\[
V^n = \underset{
\begin{tiny}\begin{array}{cc}
&i_1 +\cdots +i_k = n\\
&i_1\leq \cdots\leq i_k
\end{array}
\end{tiny}}{\oplus}
\C x_{i_1}\cdots x_{i_k}.
\]
Namely, the space $V^n$ consists of (homogeneous) polynomials of degree $n$
in our convention. So the degree is a synonym of the conformal weight\footnote{
However, since we will be dealing with the classification of conformal vectors and the notion
of conformal weight depends on the choice of conformal vector, we shall use the term degree
instead of the conformal weight.} 
with  respect to the conformal vector (\ref{eqn:conformal}).

\section{Classification of Conformal Vectors of Free Bosonic Vertex Algebra}

This section is the main contribution of the present work. We will classify all the
conformal vectors without any restriction on its degree
with respect to the standard gradation. The main result
is that the conformal vectors given in Subsection 2.4 exhaust all the conformal
vectors up to the action by inner automorphisms.

\subsection{Commutative Vectors}
We begin by introducing the notion of commutative vectors, which will become fundamental in our
consideration. Let $M$ be a vector space and $C(z)$ a field on $M$.
We say that $C(z)$ is a \textit{commutative field} if
\[
C(y)C(z) = C(z)C(y),
\]
which is equivalent to the OPE
\[
C(y)C(z)\sim 0
\]
under the assumption that $C(y)$ is local with itself.
It is also equivalent to the condition that
the Fourier modes $C_{(n)},n\in\Z$ form a representation
of commutative algebra (more precisely, of the polynomial ring)
\[
C_{(m)}C_{(n)} = C_{(n)}C_{(m)},\quad m,n\in\Z.
\]
When $M$ is a vertex algebra $V$ and we have a vector $c\in V$ such that
\[
C(z) = Y(c,z),
\]
is a commutative field, then we call such $c\in V$ a \textit{commutative vector}.
Namely, an element $c\in V$ is a commutative vector if
and only if
\[
c_{(n)}c = 0, \quad(n\geq 0).
\]
For example, the vacuum vector is clearly a commutative
vector.

For the free bosonic vertex algebra $V = \mathcal{F}_0$,
we have the following. 

\begin{theorem}[Uniqueness of commutative vectors]\label{theorem:3.1}
Let $V =\mathcal{F}_0$ be the free bosonic vertex algebra. Then a vector
$c\in V$ is a commutative vector if and only if it is a scalar multiple of the vacuum
vector.
\end{theorem}

The proof, which is based on Wick's theorem, will be given in 
Subsection 3.4.

\subsection{Classification of Heisenberg and Virasoro Vectors}

As an application of the last theorem, we will give the classification of
Heisenberg and Virasoro vectors of the free bosonic vertex algebra
$V = \mathcal{F}_0$.

Let us first consider the case of Heisenberg vectors.
By the definition, an element $h \in V$ is a Heisenberg vector if and only if 
\[
h_{(n)}h =
\begin{cases}
0&\quad (n\geq 2)\\
\vac&\quad (n=1).
\end{cases}
\]
The condition $h_{(0)}h = 0$ follows from the skew symmetry 
as in Remark \ref{remark:virasoro}.
Let us decompose $h$ into the homogeneous components according to the standard 
gradation:
\[
h = h^0 + h^1 + \cdots + h^d, \quad h^d\neq 0.
\]
Suppose $d \geq 2$. Then $h^d_{(n)}h^d$ is the 
component of $h_{(n)}h$ of degree $2d-n-1$ and
we have 
\[
h^d_{(n)}h^d = 0\quad (n\geq 0).
\]
This implies $h^d \in \C$, which is a contradiction.
Therefore, $h$ does not have a component of degree greater
than one and $h$ must be written as
\[
h = a + bx_1 \quad (a,b\in\C).
\]
Substitute this expression in the definition of Heisenberg vectors 
and compare the coefficients.
Then we see:

\begin{theorem}[Classification of Heisenberg
vectors]\label{theorem:3.3} An element of $h$ of free bosonic vertex
algebra is a Heisenberg vector if
and only if  $h$ is written as
\[
h = \mu \pm x_1,\quad\text{i.e.,}\quad Y(h,z) = \mu I(z)\pm \alpha(z)
\]
for some scalar $\mu\in \C$.
\end{theorem}

Now let us turn to the case of Virasoro vectors. The strategy is the same as above.
For a Virasoro vector $v\in V$ with homogeneous decomposition
\[
v = v^0 + v^1 +v^2 +v^3 +
\dots +v^d,\quad v^d\neq 0,
\]
suppose $d\geq 3$, then
\[
v^d_{(n)}v^d = 0\quad (n\geq 0)
\]
and $v^d$ must be proportional to $1$, thus we have 
a contradiction. So $d\leq 2$ and
\[
v = a +bx_1 +c x_1^2 +dx_2\quad
(a,b,c,d \in\C).
\]
and by substituting it into the definition of Virasoro vectors, we have a 
condition for $v$ to be a Virasoro vector:
\[
b^2 = 2a,\quad
4bc = 2b,\quad
4c^2 = 2c,\quad\text{and}\quad
4cd = 2d.
\]
Therefore,
\begin{theorem}[Classification of Virasoro vectors]
An element $v $ of free bosonic vertex algebra $V$ is
a Virasoro vector if and only if $v$ is written as
\[
v = \frac{1}{2}\mu^2 + \mu x_1 + \frac{1}{2}x_1^2 + \lambda x_2,
\]
i.e.,
\[
Y(v,z) = \frac{1}{2}\mu^2 I(z) + \mu \alpha(z) +
\frac{1}{2}\NO \alpha(z)\alpha(z)
\NO +\lambda \partial \alpha(z)
\]
for some scalar $\lambda,\mu\in \C$. 
\end{theorem}

Now for a Virasoro vector $v = \frac{1}{2}\mu^2 + \mu x_1 +\frac{1}{2}x_1^2
+\lambda x_2$, we have
\[
L_{-1}:= v_{(0)} = T,\quad 
L_0 := v_{(1)} = \mu\frac{\partial}{\partial x_1} +
\sum_{i =1}^\infty ix_i\frac{\partial}{\partial x_i}.
\]
Such $L_0$ is always semisimple and thus

\begin{corollary}[Classification of conformal vectors]
\label{corollary3.4}
All the Virasoro vectors in the free bosonic vertex algebra are conformal vectors
and are given by the theorem above.
\end{corollary}

\begin{remark}
If $\mu \neq 0$, then the gradation defined by the conformal vector
$v = \frac{1}{2}\mu^2 + \mu x_1 +\frac{1}{2}x_1^2
+\lambda x_2$ differs from the standard one.
\end{remark}

\subsection{Applications}
Let us determine the automorphism group of the free bosonic vertex
algebra 
$V = \C[x_1,x_2,\dots]$. First, since $V$ is generated by $x_1$, we have:

\begin{lemma}
An automorphism $\sigma$ of the free bosonic vertex algebra $V$ is 
uniquely determined by $\sigma(x_1)$.
\end{lemma}

Then because an automorphism $\sigma$ maps a Heisenberg vector to another, 
we have
\[
\sigma(x_1) = a\pm x_1
\]
and conversely, by the above lemma, such an automorphism is unique if
it exists. Now since $\alpha_1$ is a derivation which decreases the
degree by $1$,
\[
\tau_a = \exp{(a\alpha_1)} =
\exp{\left(a\frac{\partial}{\partial x_1}\right)},\quad a\in\C
\]
does converge and gives an inner automorphism of $V$. 
On the other hand, it is easy to see that
\[
\iota:x_1\longmapsto -x_1
\]
uniquely extends to an automorphism of $V$ which preserves the standard
gradation and is not
inner.  Since they satisfy
\[
\tau_a (x_1) = a+x_1,\quad \iota\tau_a = a-x_1,
\]
we have
\[
\Aut V = \{\tau_a,\iota\tau_a\,|\, a\in \C\}.
\]
Therefore
\[
\Inn\, V = \Inn_{(1)}\,V =\{\tau_a\,|\, a\in \C\}\cong \C,\quad 
\Autn^0\,V =\{\id,\iota\}\cong \Z/2\Z.
\]
Further we have
\[
\iota \tau_a\iota^{-1} = \tau_{-a}
\]
and $\Inn V\cap\Autn^0 V =\{\id\}$. Summarizing, we have:
\begin{theorem}
The automorphism group of the free bosonic vertex algebra is a semi--direct
product
\[
\Aut V = \Inn V \rtimes \Autn^0\, V
\]
where
$
\Inn V = \{\tau_a\,|\, a\in \C\}\cong \C
$
and
$
\Autn^0\, V = \{\id,\iota\}\cong \Z/2\Z.
$
\end{theorem}

Now by the classification given in Subsection 3.2, a conformal vector is written
as
\[
v = \frac{1}{2}\mu^2 +\mu x_1 +\frac{1}{2}x_1^2 + \lambda x_2
\]
for some scalars $\lambda,\mu\in\C$. An inner automorphism $\tau_a, a\in \C$
maps $v$ to a vector of degree $2$ if and only if $a = -\mu$
and in this case
\[
\sigma(v) = \frac{1}{2}x_1^2 +\lambda x_2.
\]
This is a conformal vector described in Subsection 2.4 and we have:

\begin{corollary}
Any conformal vector of the free bosonic vertex algebra is transformed by an inner
automorphism to a unique conformal vector of the form
\[
v = \frac{1}{2}x_1^2 + \lambda x_2
\]
for some $\lambda\in \C$
\end{corollary}

Finally let us mention about the complete reducibility
of the vertex algebra $V$ as a representation of $sl_2(\C)$ through
the homomorphism
\[
\begin{array}{ccc}
sl_2(\C)&\longrightarrow&\Vir\\
E&\longmapsto&-L_{1}\\
H&\longmapsto&-2L_0\\
F&\longmapsto&L_{-1}
\end{array}
\]
where $\{E,H,F\}$ is the standard basis of $sl_2(\C)$ and the $\Vir$
module structure on $V$ is given by a conformal vector $v$.
Then the weight space decomposition is
\[
V =\oplus_{k\in\N}V_{-2k},\quad
V_{-2k} = \{a\in V\,|\,Ha = -2ka\}.
\]
By applying an inner automorphism, we may assume that the conformal vector is
written as
\[
v = \frac{1}{2}x_1^2 + \lambda x_2.
\]
Then the gradation is the standard one and in particular,
\[
V_0 =\C 1,\quad V_{-2} = \C x_1.
\]
Therefore, if $L_1 x_1 = 0$, then we have a direct sum decomposition
\[
V = V_0\oplus\left(
\oplus_{k>0}V_{-2k}\right)
\]
and $V$ is completely reducible since the latter factor is a direct sum of Verma
modules with negative highest weights. Otherwise, the submodule generated
by $x_1$ is isomorphic to $M^*(0)$, the dual Verma module with highest weight $0$,
and $V$ cannot be completely reducible. Therefore $V$ is completely reducible
if and only if $L_1x_1 = 0$, namely, $v =\frac{1}{2}x_1^2$.

\begin{proposition}
A conformal vector $v$ gives a completely reducible $sl_2$--module structure on
$V$ if and only if $v$ is the image of the conformal vector 
$\frac{1}{2}x_1^2$ by an inner
automorphism.
\end{proposition}

Since $\frac{1}{2}x_1^2$ is the unique conformal vector which is fixed 
by involution $\iota$, we have the following observation.

\begin{corollary}
A conformal vector $v$ of the free bosonic vertex algebra $V$ gives a completely
reducible $sl_2$--module structure on $V$ if and only if $v$ is fixed by some
non--trivial automorphism of $V$.
\end{corollary}
\subsection{Proof of Theorem \ref{theorem:3.1}}

Before proceeding to the proof, let us prepare the
notion of partitions. A partition is a sequence of 
non--negative integers
\[
\lambda = (\lambda_1,\lambda _2,\dots)
\]
such that
\[
\lambda_1 \geq \lambda _2\geq\dots \quad\text{and}\quad \lambda _n = 0
\quad \text{for sufficiently large}\quad n.
\]
Let $P$ denotes the set of all partitions. Set
\[
\ell(\lambda ) = \max \{n|\,\lambda _n \neq 0\},
\quad |\lambda| = \lambda _1 + \lambda _2 +\cdots +\lambda_{\ell(\lambda)}.
\]
They are called the length and the size of $\lambda$ respectively.
We denote by $\phi$ the partition with all the entries being zero.
Then 
$|\phi|= \ell(\phi)= 0$. Let us denote $P_\ell$ the set of partitions of length $\ell$.

Let $V = \C[x_1,x_2,\dots]$ be the free bosonic vertex algebra.
For a partition $\lambda = (\lambda_1,\lambda_2,\dots)$, set
\[
x_\lambda = x_{\lambda_1}x_{\lambda_2}\cdots x_{\lambda_{\ell(\lambda)}},
\quad x_\phi = 1.
\]
Then the set of $P$ of partitions indexes a basis of $V$:
\[
V = \oplus_{\lambda\in P}\C x_{\lambda}.
\]

Now, let $c$ be a commutative vector and
\[
c = \sum_{\lambda\in P}c_\lambda x_\lambda\quad (c_\lambda\in \C)
\]
be its representation by the basis. Then to prove Theorem 3.1, i.e.,
to show that $c$ is proportional to $1$, it suffices to see that
\[
c_\lambda = 0\quad (\lambda \neq 0).
\]
For the purpose, decompose $c$ according to the length and take
the longest part of them:
\[
\sum_{\lambda\in P,\ell(\lambda) = \ell}c_\lambda x_\lambda
\]
where $\ell = \max \{\ell(\lambda)|\,c_\lambda \neq 0\}$. 
Introduce the total order defined on the set $P_\ell$ by
\[
\lambda> \mu\quad \text{if}\quad
\lambda_1 = \mu_1,\dots,\lambda_{i-1} = \mu_{i-1},\lambda_{i} > \mu_{i}
\quad \text{for some}\quad i
\]
and take the greatest one $\nu$ in the set
$\{\lambda\in P_\ell|\,c_\lambda \neq 0\}$. Now suppose $\ell\geq 1$
and let us derive a contradiction.

By Wick's theorem, we have
\[
\begin{split}
Y(c,y)Y(c,z)&\sim \sum_{|\lambda|=|\mu|=\ell, \lambda,\mu\leq \nu}
c_\lambda c_\mu
 \langle \partial^{(\lambda_i-1)}\alpha(y)\partial^{(\mu_j-1)}\alpha(z)\rangle\times\\
&\quad\times\NO Y(x_{\lambda_1}\cdots \hat{x}_{\lambda_i}\cdots
x_{\lambda_\ell},y) Y(x_{\mu_1}\cdots \hat{x}_{\mu_j}\cdots x_{\mu_\ell},z)\NO \\
&+\text{(shorter lenght terms)}.
\end{split}
\]
In the summation of the longest terms, the ones with the highest possible
pole at $y = z$ is
\[
\sum_{
\begin{tiny}
\begin{array}{c}
|\lambda|=|\mu|=\ell,\\
\lambda,\mu\leq \nu,\\
\lambda_1 = \mu_1 = \nu_1
\end{array}
\end{tiny}
}
c_\lambda c_\mu K_{\lambda\mu}\frac{1}{(y-z)^{\lambda_1 +\mu_1}}
Y(x_{\lambda_2}\cdots x_{\lambda_\ell}x_{\mu_2}\cdots x_{\mu_\ell},z)
\]
where $K_{\lambda\mu}$ is a non--zero scalar. Further the greatest term in this
summation is
\[
K_{\nu\nu}c_{\nu}^2 
\frac{1}{(y-z)^{2\nu_1}}
Y((x_{\nu_2})^2\cdots(x_{\nu_\ell})^2,z).
\]
By the assumption that $Y(c,y)Y(c,z)\sim 0$, we must have
\[
c_{\nu}^2  = 0.
\]
This contradicts the choice of $\nu$ and
we conclude that $\ell = 0$. Therefore a commutative vector $c\in V$ must be
a scalar in $V = \C[x_1,x_2,\dots]$.

\section{Discussion}
In this paper, we have classified Heisenberg vectors, Virasoro vectors, conformal vectors and
automorphisms of the free bosonic vertex algebra. All the results obtained here depend on the
fact that a commutative vector of the vertex algebra is proportional to the vacuum vector (Theorem
3.1). Indeed, this reduced the classification of Heisenberg and Virasoro vectors to those of
degree less than or equal to one and two respectively.

However, such uniqueness of commutative vectors is no more true for general vertex algebra. For
example, let $L$ be an integral lattice and $V_L$ be the vertex algebra associated
with
$L$. Then, for any $\gamma\in L$ such that $(\gamma|\gamma)\geq 0$, the vector $1\otimes e^\gamma$
is a commutative vector.

Correspondingly, the failure of the uniqueness of commutative vectors complicates the
classification of Heisenberg vectors, Virasoro vectors and so on.
In fact, the vertex algebra $V_L$ for a positive definite $L$, for example, has a conformal
vector having arbitrarily high degree with respect to the standard gradation: let
$v$ be a conformal vector of degree $2$ and take $\gamma\in L$ such that $(\gamma,\gamma)\geq 4$.
Then the vector $v + T(1\otimes e^\gamma)$ is a conformal vector with
degree
$\frac{1}{2}(\gamma,\gamma) + 1$, which can be arbitrarily large. Moreover, this vector is
obtained from $v$ by the inner automorphism $\exp{(\kappa (1\otimes e^\gamma)_{(0)})}$
where $\kappa = 2/(2-(\gamma|\gamma))$ which does
converge. In particular, this shows that the group of inner automorphisms is quite large in this
case.

The last construction is generalized as follows. Let $v$ be a conformal vector of a vertex
algebra and $u$ be a primary vector with conformal weight $\Delta$ with respect to
$v$, i.e.,
\[
v_{(1)}u = \Delta u,\quad \text{and}\quad v_{(n)}u = 0, \quad (n\geq 2),
\]
such that $u$ is a commutative vector. Then the vector $v + Tu$ is always a Virasoro vector, and if
$\Delta
\neq 1$, then it is obtained from
$v$ by the inner automorphism
$\exp{(\kappa u_{(0)})}, \kappa = 1/(1-\Delta)$ if it makes sense.

These examples illustrate the complicated situation for general vertex algebras.

\subsection*{Acknowledgment}
One of the authors (K.N) would like to express his deep
gratitude to professor V.G.
Kac for giving him an opportunity to stay at MIT and heading his concern to
vertex algebras as well as his continuous encouragement and advises. He also
thanks professor K. Harada and M. Miyamoto for their interests and valuable discussions.
Another author would like to thank professor I. Frenkel for advising him to learn
the theory of vertex operator algebras and Moonshine several years ago.


\newpage
\begin{thebibliography}{100}
\bibitem[B1]{B86}
Borcherds, R.: Vertex algebras, Kac--Moody algebras, and the Monster, Proc. Natl. Acad. Sci. USA,
\textbf{83}, 3068--3071(1986)
\bibitem[B2]{B92}
Borcherds, R.: Monstrous moonshine and monstrous Lie superalgebras. Invent. Math.,
\textbf{109}, 405--444(1992)
\bibitem[BPZ]{BPZ} Belavin, A., Polyakov, A. and Zamolodchikov:
Infinite conformal symmetries in two--dimensional quantum field theory.
Nucl. Phys., \textbf{B241}, 333-380(1984)
\bibitem[DL]{Dong--Lepowsky}
Dong, C. and Lepowsky, J.: Generalized Vertex Algebras and Relative
Vertex Operators, Progress in Mathematics, \textbf{112}, Birkh\"{a}user (1993)
\bibitem[FHL]{FHL}
Frenkel, I.B., Huang, Y. and Lepowsky, J.:
On axiomatic approach to vertex algebras to vertex operator algebras and modules,
Mem. Amer. Math. Soc.,\textbf{104} No.494(1993)
\bibitem[FKRW]{FKRW}Frenkel, E., Kac,V., Radul, A., Wang,W.:$W_{1+\infty}$
and $W(gl_{\infty})$ with central charge $N$. Commun. Math. Phys.,
\textbf{170}, 337--357(1995)
\bibitem[FLM]{FLM}
Frenkel, I.B., Lepowsky, J., Meurman, A.: Vertex operator algebras and the Monster,
Academic Press, (1988)
\bibitem[G]{G89}
Goddard, P.: Meromorphic conformal field theory. In V. G. Kac, editor,
Infinite--dimensional Lie algebras and groups, Adv. Ser. in Math. Phys., \textbf{7},
World Scientific, 556-587(1989)
\bibitem[K]{K96}
Kac, V.G.: Vertex algebras for beginners, University Lecture series,  \textbf{10},
AMS.(1996)
\bibitem[Li]{Li}
Li, H.-S.: Local systems of vertex operators, vertex superalgebras and
modules, J. Pure Appl. Algebra, \textbf{109}, 143--195 (1996). 
Local systems of twisted vertex operators, vertex operator superalgebras and twisted
modules. Moonshine, the Monster, and related topics (South Hadley, MA, 1994),
203--236,
Contemp. Math., \textbf{193},AMS.(1996)
\bibitem[Lian]{Lian}
Lian, B. H.: On the classification of simple vertex operator algebras, Commun. Math. Phys.
\textbf{163}, 307--357(1994)
\bibitem[LZ]{LZ}
Lian, B. H. and Zuckerman, G.J.: Commutative quantum operator algebras. J. Pure Appl. Algebra,
\textbf{100}, 117--140(1995)
\bibitem[MN]{Matsuo--Nagatomo}
Matsuo, A. and Nagatomo, K.: On axioms of vertex algebras and locality of 
quantum fields, in preparation.
\end{thebibliography}
\end{document}